\documentclass[twocolumn,aps]{revtex4}

\usepackage{graphicx}
\usepackage{dcolumn}
\usepackage{bm}
\begin{document}
\draft
\title{
Measurement of the photonic de Broglie wavelength of biphotons 
generated by spontaneous parametric down-conversion
}

\author{Keiichi Edamatsu, Ryosuke Shimizu, and Tadashi Itoh} 
\address{
Division of Materials Physics, Graduate School of Engineering Science, 
Osaka University, Toyonaka 560-8531, Japan
}

\date{\hspace*{12em}}
\begin{abstract}
Using a basic Mach-Zehnder interferometer, 
we demonstrate experimentally the measurement of the
photonic de Broglie wavelength of an entangled photon pair 
(a biphoton) generated by spontaneous parametric down-conversion. 
The observed interference manifests the 
concept of the photonic de Broglie wavelength.
The result also provides a proof-of-principle of the
quantum lithography that utilizes 
the reduced interferometric wavelength.
\end{abstract}
\vspace{5mm}


\maketitle

\narrowtext
%
%
The nature of entanglement or quantum correlation between two or more 
particles has attracted great interest and produced many 
applications in quantum information processing \cite{Qinfo}, 
such as quantum computation \cite{Qcomp}, 
quantum cryptography \cite{Qcrypt}, 
and quantum teleportation \cite{Bouwmeester,Furusawa}.
Especially, 
a number of novel experiments have used
parametric down-converted photons
because they produce a superior environment for the realization
of ideas concerning quantum entanglement.
In addition to such applications to quantum informatics,
the genuine quantum optical properties of parametric down-converted
photons are also very attractive.
Jacobson {\it et al.}~\cite{Jacobson} proposed the concept of
the ``photonic de Broglie wavelength'' in multiphoton states.
They argued that the photonic de Broglie wavelength of an ensemble of photons 
with wavelength $\lambda$ and number of photons $N$ 
can be measured to be $\lambda/N$ using a special interferometer 
with ``effective beam splitters'' that do not split the multiphoton states 
into constituent photons. 
Following this concept,
Fonseca {\it et al.}~\cite{Fonseca} 
used a kind of Young's double slit interferometer to
measure the photonic de Broglie wavelength of entangled photon pairs (``biphotons'') generated by parametric down-conversion. 
In the concept of photonic de Broglie wavelength,
the interference and diffraction properties of multiphoton states
are governed by their reduced wavelength.
Based on this idea, 
Boto {\it et al.}~\cite{Boto} proposed the principle of ``quantum lithography'',
utilizing the reduced interferometric wavelength of 
non-classical $N$ photon states
for optical imaging beyond the classical diffraction limit.
Recently, a proof-of-principle experiment in quantum lithography
was demonstrated by D'Angelo {\it et al.}~\cite{DAngelo}
utilizing parametric down-converted biphotons.
In this letter, we propose and demonstrate the 
measurement of the photonic de Broglie wavelength for $N$=2 state
in a very simple and straightforward manner,
utilizing biphotons generated by parametric down-conversion 
and a basic Mach-Zehnder (M-Z) interferometer.
We show that not only the ``wavelength'' but also the coherence length
of the biphoton are different from those of a single photon.
We also discuss the nature of biphoton interference,
which is essentially governed by the quantum correlation
between two constituent photons. 

\begin{figure}
\includegraphics[width=80mm]{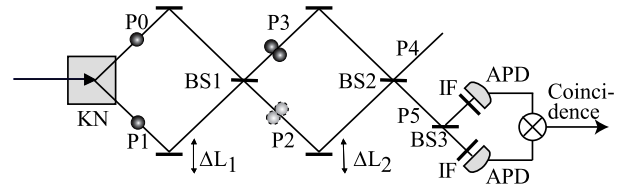}
\caption{
Schematic experimental setup of the biphoton interference. KN: KNbO$_3$ crystal, BS1$\sim$3: beam splitters, IF: interference filters, APD: avalanche photodiodes.
}
\label{fig-exp}
\end{figure}

\begin{figure}
\includegraphics[width=70mm]{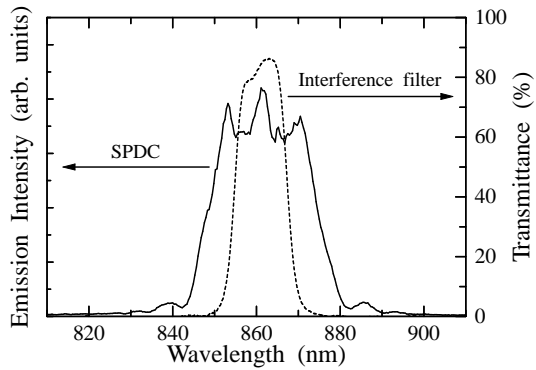}
\caption{
Solid curve: spectrum of the signal light generated by spontaneous parametric 
down-conversion; 
Dashed curve: transmission spectrum of the interference filter 
(IF in Fig.~\ref{fig-exp}) placed in front of the detectors
used for the interference measurements.
}
\label{fig-spectr}
\end{figure}

\begin{figure}
\includegraphics[width=65mm]{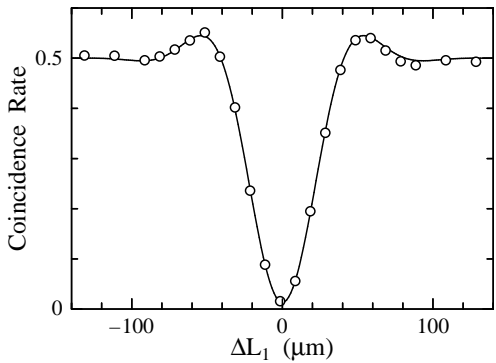}
\caption{
Coincidence-photon-counting rate detected at the two output ports
of BS1 as a function of the optical path-length difference ($\Delta L_1$)
between the two paths from KN and BS1 in Fig.~\ref{fig-exp}. 
Open circles indicate experimental data, and the solid curve is a fitted function
that assumes the observed photons have a rectangular spectral shape.
}
\label{fig-tpi}
\end{figure}

The schematic view of the apparatus used in our experiment 
is shown in Fig.~\ref{fig-exp}. 
Pairs of photons were generated by spontaneous parametric 
down-conversion (SPDC) in a 5-mm-long KNbO$_3$ (KN) crystal,
the temperature of which is controlled and stabilized to
within 0.01$^\circ$C. 
The pump source of the SPDC was the second harmonic light
of a single longitudinal mode Ti:sapphire laser 
operating at $\lambda_0$=861.6 nm (linewidth $\Delta\nu_0\sim 40$ MHz). 
We selected correlated photon pairs traveling along two paths (P0 and P1)
by two symmetrically placed pinholes.
The central wavelength of the photon pair was tuned to 
degenerate at 861.6 nm by controlling the KN temperature.
Figure~\ref{fig-spectr} shows the emission spectrum of SPDC
after the pinhole, 
together with the transmission spectrum of the interference filters
used for the measurement (see below).
The M-Z interferometer was composed of two 50/50\% beam splitters 
(BS1 and BS2).
A biphoton was generated at either one (P2 or P3) of the interferometer arms
when a pair of down-converted photons simultaneously entered 
at both input ports of a beam splitter (BS1), 
as a result of Hong-Ou-Mandel (HOM) interference \cite{Hong}.
By observing the HOM interference,
$\Delta L_1$ was adjusted so that the coincidence rate 
detected at both output ports (P2 and P3) of BS1 was minimized
(See Fig.~\ref{fig-tpi}).
Thus we can prepare the entangled biphoton state
\begin{eqnarray}
 \frac{1}{\sqrt2}\left(|2\rangle_{\rm P2} |0\rangle_{\rm P3} 
 + |0\rangle_{\rm P2} |2\rangle_{\rm P3} \right)
\label{eq0}
\end{eqnarray}
in the M-Z interferometer, 
where $|N\rangle_{i}$ denotes $N$ photons travel along the arm $i$.
The path-length difference ($\Delta L_2$) between the two arms of the M-Z interferometer was controlled by a piezoelectric positioner,
which was capable of controlling $\Delta L_2$ with nanometer resolution.
The two paths were combined at the output beam splitter (BS2), 
and the biphoton interference was measured
at one of the output ports (P5) of BS2.
The biphoton interference pattern was recorded by a two-photon detector, consisting of a 50/50\% beam splitter (BS3) and two avalanche photodiodes (APD) followed by a coincidence counter.
We put an interference filter 
(IF: center wavelength $\lambda_c$ = 860 nm, 
bandwidth $\Delta\lambda$ = 10 nm; See Fig.~\ref{fig-spectr})
in front of each APD to eliminate undesired background light.
In general, the interferometer was designed to measure the photonic de Broglie wavelength of the biphoton without using any special 
``effective beam splitters''. 
For comparison, we also measured the usual single-photon interference 
by using a single detector and blocking one of the input ports.

It is worth discussing the interference patterns
we expected to obtain in our experiment.
For simplicity, we consider only degenerate single-frequency photons. 
The single-frequency treatment is adequate for predicting 
most distinct properties of interference patterns for both
one-photon and two-photon detection,
although it will be necessary to consider multi-frequency treatment 
in order to discuss
more-detailed phenomena such as the coherence length of the
interference.
The one- and two-photon counting rates ($R_5$ and $R_{55}$, respectively) 
at an output port P5 of the interferometer are described by
\begin{eqnarray}
R_5(|\psi_0,\psi_1\rangle) &=& \langle\psi_0,\psi_1|\hat{a}_5^\dag\hat{a}_5|\psi_0,\psi_1\rangle , 
\label{eq1}\\
R_{55}(|\psi_0,\psi_1\rangle) &=& \langle\psi_0,\psi_1|\hat{a}_5^\dag\hat{a}_5^\dag\hat{a}_5\hat{a}_5|\psi_0,\psi_1\rangle , 
\label{eq2}
\end{eqnarray}
where $\hat{a}_5^\dag$ and $\hat{a}_5$ are photon creation and annihilation
operators at the output port P5, 
$\psi_0$ and $\psi_1$ denote the quantum states of light at the two 
input ports P0 and P1, respectively. 
Also, the coincidence counting rate ($R_{45}$) detected
at both output ports of the interferometer is
\begin{eqnarray}
R_{45}(|\psi_0,\psi_1\rangle) = \langle\psi_0,\psi_1|\hat{a}_4^\dag\hat{a}_5^\dag\hat{a}_5\hat{a}_4|\psi_0,\psi_1\rangle . 
\label{eq3}
\end{eqnarray}
where $\hat{a}_4^\dag$ and $\hat{a}_4$ are photon creation and annihilation
operators, respectively, at the output port P4. 
The photon operators at the output ports are 
connected to those of the input ports 
through the scattering matrices of the beam splitters
and the optical path-length difference \cite{Mandel}: 
\begin{eqnarray}
\left(\begin{array}{c} \hat{a}_4\\ \hat{a}_5 \end{array} \right)
=
\frac{1}{2}\left(\begin{array}{cc} 1 & i \\ i & 1 \end{array} \right)
\left(\begin{array}{cc} e^{i\phi_2} & 0 \\ 0 & e^{i\phi_3} \end{array} \right)
\left(\begin{array}{cc} 1 & i \\ i & 1 \end{array} \right)
\left(\begin{array}{c} \hat{a}_0\\ \hat{a}_1 \end{array} \right)
\label{eq4}
\end{eqnarray}
and their Hermitian conjugates,
where $\phi_2$ and $\phi_3$ are phase retardations along the 
interferometer arms P2 and P3, respectively.
Here, we assume the M-Z interferometer consists of 
lossless 50/50\% beam splitters.
Using (\ref{eq4}),
we can calculate the counting rates (\ref{eq1})-(\ref{eq3}) for
arbitrary input states of light.
The resultant two-photon interference patterns for the case of
$|\psi_0,\psi_1\rangle$=$|1,1\rangle$, 
i.e., where both inputs are $N$=1 Fock states, are
\begin{eqnarray}
R_{55}(|1,1\rangle) &=& \frac12(1-\cos 2\phi) ,   
\label{eq5a}\\
R_{45}(|1,1\rangle) &=& \frac12(1+\cos 2\phi) ,
\label{eq5b}
\end{eqnarray}
where $\phi\equiv\phi_2-\phi_3=2\pi\Delta L_2/\lambda$ 
is the optical phase difference between the two arms, 
$\Delta L_2$ the path-length difference,
and $\lambda$ the wavelength of the input light.
Here, the constant coefficients 
of the right sides have been ommitted.
The corresponding one-photon interference for the case of
$|\psi_0,\psi_1\rangle$=$|0,1\rangle$ becomes 
\begin{eqnarray}
R_5(|0,1\rangle) &=& \frac12(1-\cos \phi) .
\label{eq6}
\end{eqnarray}
From Eqs.~(\ref{eq5a})--(\ref{eq6}),
we see that both $R_{55}$ and $R_{45}$
will have the oscillation period $\lambda/2$,
while $R_5$ has the period $\lambda$.
This oscillation period $\lambda/2$ for the two-photon counting rate $R_{55}$
is attributable to the photonic de Broglie wavelength $\lambda/N$ for the biphoton ($N$=2) state.
Although $R_{45}$ will also have the oscillation period $\lambda/2$,
it is not attributable to the photonic de Broglie wavelength
of the biphoton
because in this case the two photons are split from each other by BS2.
Furthermore, the oscillation period $\lambda/2$ in $R_{45}$
could be observed even for classical states,
while that in $R_{55}$ could not.
In fact, for the coherent input state $|0,\alpha\rangle$,
\begin{eqnarray}
R_{55}(|0,\alpha\rangle) =  R_5^2 &=& \frac{|\alpha|^4}{4}(1-\cos \phi)^2 ,   
\label{eq7a}\\
R_{45}(|0,\alpha\rangle) =  R_4R_5 
&=& \frac{|\alpha|^4}{4}(1+\cos \phi)(1-\cos \phi) \nonumber\\
&=& \frac{|\alpha|^4}{8}(1-\cos 2\phi) .
\label{eq7b}
\end{eqnarray}
Although the coherent state is classified as a classical state of light,
the oscillation period of $R_{45}(|\alpha,0\rangle)$ becomes $\lambda/2$.
Consequently, the oscillation period $\lambda/2$ of $R_{45}$
does not necessarily originate from the quantum nature of light. 
On the other hand,
in the classical treatment,
$R_{55}$ should have the same period as $R_5$
since $R_{55}$=$R_5^2$ and $R_5$$\ge$0.
Thus, the oscillation period $\lambda/2$ of $R_{55}(|1,1\rangle)$ 
reflects the quantum nature of the input field and
consistent with the photonic de Broglie wavelength
of the biphoton.
Hence, the intent of our experiment is to measure the interference pattern of 
$R_{55}(|1,1\rangle)$.

Before observing the interference pattern of the M-Z interferometer,
we measured the HOM interference to check the position of the zero path-length
difference $\Delta L_1$ and to ensure that 
the biphotons were generated at either arm of the interferometer.
The observed HOM interference, i.e., 
the coincidence counting rate between the two output ports (P2 and P3)
of BS1 as a function of the optical path-length difference ($\Delta L_1$),
is presented in Fig.~\ref{fig-tpi}.
The visibility of the HOM interference was 0.97, 
guaranteeing that the photon pair was traveling together almost perfectly 
along either arm of the interferometer when $\Delta L_1=0$.
To our knowledge, this is one of the best visibilities ever obtained 
in a HOM interference experiment.
After the measurement of the HOM interference,
$\Delta L_1$ was fixed at 0, where the coincidence rate was minimized.
Thus we prepared the entangled biphoton state (\ref{eq0})
in our interferometer.

\begin{figure}
\includegraphics[width=65mm]{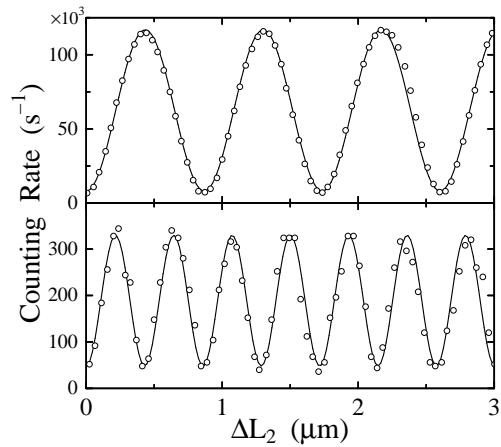}
\caption{
Interference patterns in the one-photon (upper) and two-photon (lower) 
counting rates at a path-length difference of around 0 $\mu$m.
}
\label{fig-bpi0}
\end{figure}

Figure~\ref{fig-bpi0} shows the measured interference pattern for both 
one-photon ($R_{5}(|0,1\rangle)$: upper graph) 
and two-photon ($R_{55}(|1,1\rangle)$: lower graph) detection, 
as a function of path-length difference ($\Delta L_2$) 
around $\Delta L_2$$\sim$0 $\mu$m. 
Note that one of the input ports of the interferometer 
was blocked during measurement of the one-photon
counting rate; otherwise, no interference was expected.
For both one- and two-photon counting rates,
clear interference fringes were observed.
The fringe visibilities of the one- and two-photon interferences 
were 0.88 and 0.75, respectively.
It can be seen in the figure
that the one-photon interference has a period of 
approximately 860 nm, 
whereas the period in the two-photon interference 
is approximately 430 nm. 
This result clearly indicates that the biphoton state 
exhibits the interference as a ``wave'' whose length 
is half that of the one-photon state. 
This is consistent with the prediction that the photonic de Broglie 
wavelength of the biphoton state would be $\lambda_c/2$.
Thus, we have clearly measured the photonic de Broglie wavelength 
of the biphoton generated by SPDC. 

\begin{figure}
\includegraphics[width=65mm]{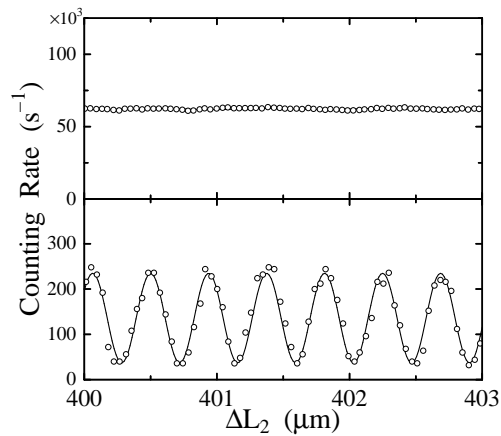}
\caption{
Interference patterns in the one-photon (upper) and two-photon (lower) 
counting rates at a path-length difference of around 400 $\mu$m.
}
\label{fig-bpi400}
\end{figure}

We have also observed the difference in the coherence length 
between the one- and two-photon counting rates, as demonstrated 
in Fig.~\ref{fig-bpi400}. 
Although the interference oscillation in the one-photon counting 
rate disappears at $\Delta L_2$$\sim$400 $\mu$m, the interference 
of the two-photon counting rate remains until the path-length difference is
much larger, indicating that biphotons have much longer 
coherence lengths than do single photons.
Since the spontaneous parametric down-converted photons have considerably
wide spectral widths, the coherence length of the one-photon counting rate
is governed by the spectral bandwidth $\Delta\lambda$ 
of the interference filters placed in front of the detectors 
(See Fig.~\ref{fig-spectr}).
Thus, the coherence length of the one-photon counting rate becomes
very short ($\lambda_c^2/\Delta\lambda\sim 70$ $\mu$m).
On the other hand, the coherence length of the two-photon counting rate
is governed by the spectral width of the sum frequency of signal 
($\nu_s$) and idler ($\nu_i$) photons. 
This sum frequency is identical to the frequency of pump photons (2$\nu_0$)
of the SPDC.
Since we used the second harmonic light of the single longitudinal mode 
continuous laser as a pump source, its coherence length is very
long ($c/\Delta\nu_0\sim 400$ cm).
As a result, a clear interference fringe was observed for the two-photon
counting rate even at $\Delta L_2$$\sim$400 $\mu$m \cite{coherence}, 
whereas almost no fringe was observed for the one-photon counting rate.
This is the direct consequence of the
frequency correlation:
\begin{eqnarray} 
 \nu_s + \nu_i = 2\nu_0
\label{eq8}
\end{eqnarray} 
between the constituent signal and idler photons of the biphoton.
Therefore, the fringe interval and coherence length of the two-photon counting
rate consistently indicate that the biphoton is associated with the
photonic de Broglie wavelength:
\begin{eqnarray} 
\lambda_b = \frac{c}{\nu_s + \nu_i} 
          = \frac{c}{2\nu_0} 
          = \frac{\lambda_0}{2}, 
\label{eq9}
\end{eqnarray} 
where the quantum correlation (\ref{eq8}) 
between the two photons plays an important role.

Thus far, a number of works concerning two-photon 
interference have used parametric down-converted photons and either 
Mach-Zehnder or Michelson interferometer \cite{Rarity,Ou,Brendel,Shih}. 
However, those previous experiments did not intend to observe the 
photonic de Broglie wave. 
Most of them \cite{Rarity,Ou,Shih} 
detected two split photons at each of the output ports of the interferometer.
In our experiment, 
by detecting the two-photon counting rate at one of the output ports, 
we directly showed that the observed biphoton interference manifests 
the concept of the photonic de Broglie wavelength.
It is also noteworthy that 
our result provides a direct proof-of-principle of the idea of
quantum lithography \cite{Boto},
where the reduced interferometric wavelength is utilized to improve
the spatial resolution of optical imaging beyond the classical 
diffraction limit.
Finally, we note the relationship between our experiment and
the non-local nature of the entangled photon pairs, i.e., biphotons,
generated by parametric down-conversion or atomic cascade fluorescence.
As previously proposed \cite{Franson} and demonstrated \cite{Ou2,Kwiat}, 
two-photon quantum interference occurs for biphotons
even using two spatially separated interferometers.
With these results taken together, we can understand that
the interferometric properties of the biphoton originate from
its non-local quantum correlation in frequency 
between the constituent photons,
but not from the spatial closeness of the two photons.
The concept of photonic de Broglie wavelength is not inconsistent
with the standard quantum treatment of light;
rather, it provides an intuitive and essential way 
to understand the interferometric properties of 
the entangled multiphoton states.

In conclusion,
we have successfully measured the photonic de Broglie wavelength of 
the biphotons generated by spontaneous parametric down-conversion 
utilizing a  Mach-Zehnder interferometer. 
We showed that
the nature of biphoton interference is governed essentially by
the quantum correlation in frequency between the two constituent photons. 
Our results will encourage novel applications, 
such as quantum lithography, 
that utilize quantum-mechanical interference of entangled photons.

This work was supported by 
a Grant-in-Aid for Scientific Research from the Ministry of Education, 
Science, Sports and Culture of Japan,
and by the program ``Research and Development 
on Quantum Communication Technology'' of the Ministry of Public
Management, Home Affairs, Posts and Telecommunications of Japan.


\begin{references}
\bibitem{coherence}
A slight decrease in the fringe amplitude of the two-photon interference might
originate from partial misalignment of the interferometer
at a long path-length difference.
\end{references}
\end{document}